\documentclass{revtex4}
\usepackage{url}
\usepackage{xcolor}
\usepackage{hyperref}
\colorlet{shadecolor}{gray!15}
\definecolor{greenLinks}{rgb}{0,0.6,0}
\definecolor{blueLinks}{rgb}{0,0,0.6}
\definecolor{redLinks}{rgb}{0.6,0,0}
\definecolor{tempText}{rgb}{0.55,0.10,0.67}
\definecolor{eprintLinks}{rgb}{0.4,0.4,0.4}
\definecolor{journalLinks}{rgb}{0.6,0,0}
\usepackage{amssymb}
\usepackage{amsmath,color}
\usepackage{epsfig}
\usepackage{graphicx,epsf,epsfig}
\usepackage{footnote}
\usepackage{multirow}
\usepackage{graphicx}
\usepackage{subfigure}
\usepackage{lipsum}
\usepackage{multirow} 
\usepackage{tabulary}
\usepackage{rotating}
\usepackage{array}
\usepackage{color}

\begin{document}

\title{Higgs sector with spontaneous CP violation in $S(3)$ Standard Model}

\allowdisplaybreaks \allowdisplaybreaks[2]
 \newcommand{\AddrFCFMBUAP}{
  Fac. de Cs. F\'{\i}sico Matem\'aticas, 
  Benem\'erita Universidad Aut\'onoma de Puebla,\\
  Apdo. Postal 1152, Puebla, Pue.  72000, M\'exico.}
 \newcommand{\AddrFCEBUAP}{
  Fac. de Cs. de la Electr\'onica, 
  Benem\'erita Universidad Aut\'onoma de Puebla,\\
  Apdo. Postal 542, Puebla, Pue. 72000, M\'exico.}
\newcommand{\AddrUdeS}{
 Departamento de F\'{\i}sica, Universidad de Sonora, Apdo. Postal 1626,
 Hermosillo, Son.  83000, M\'exico.}
 
 \author{E. Barradas-Guevara}
 \email{barradas@fcfm.buap.mx}
 \affiliation{\AddrFCFMBUAP}
%
%
\author{O. F\'elix-Beltr\'an}
 \email{olga.felix@correo.buap.mx}
 \affiliation{\AddrFCEBUAP}
%
%
\author{E. Rodr\'{\i}guez-J\'auregui}
 \email{ezequiel.rodriguez@correo.fisica.uson.mx}
 \affiliation{\AddrUdeS}
%

\begin{abstract}
Conditions for spontaneous Charge-Parity (CP) violation in the scalar potential sector of general $S(3)$ Higgs-doublet model
(3HDM) are analyzed. An analysis of  the Higgs sector of the minimal $S(3)$-invariant extension of the Standard Model including CP violation arising from the spontaneous breaking of the electroweak symmetry is presented. This extended Higgs sector with three SU(2) doublets Higgs fields with complex vev's provides an interesting scenario to analyze the Higgs masses spectrum, trilinear self-couplings and CP violation. We present how the spontaneous electroweak symmetry breaking, coming from three $S(3)$ Higgs fields, gives an interesting scenario with nine physical Higgs  and three Goldstone bosons, when spontaneous CP violation arises from the Higgs field $S(3)$ singlet $H_S$. 
\end{abstract}

\maketitle

\bigskip

\noindent {\bf Keywords:} Higgs bosons; CP violation; Non Abelian permutational symmetry $S(3)$; 3HDM.

\section{INTRODUCTION}
\noindent It is well known that discrete symmetries,  such as charge inversion  (C), parity (P) and time reversal (T), are useful in understanding elementary particle interactions.  These symmetries are  preserved in strong and electromagnetic processes~\cite{Glashow:1976nt,Glashow:1961tr,Weinberg:1967tq}, but violated in weak interactions as it has been proved since the discovery of CP violation in neutral kaon decays in 1964~\cite{bib1}.

The Standard Model (SM) of electroweak physics is an SU(2)$_L\times$ U(1)$_Y$ gauge theory coupled to quarks, leptons and one complex hypercharge-one, SU(2)$_L$ doublet of scalar fields~\cite{Weinberg:1967tq, Higgs:1964pj}. As it is known in the SM there are three quark families, as first proposed in the theory of CP violation by M. Kobayashi and T. Maskawa~\cite{Kobayashi:1973fv},  based on phases in the coupling constants describing the weak charge-changing transitions of quarks, CP-violating effects  arise when there are at least three quark families. For three quark families these couplings are described by the unitary
$3\times3$ Cabibbo-Kobayashi-Maskawa (CKM) matrix~\cite{Kobayashi:1973fv,Cabibbo:1963yz}. With three quark families, one cannot make a phase redefinition of quarks so that all the couplings are real. To return to the CKM matrix, we have that a unitary $3\times 3$  matrix can be parameterized with three mixing angles $\theta_{ij}$ and one CP violating phase $\delta$  in the standard parameterization as suggested  by M. Kobayashi and T. Maskawa~\cite{bib3};
with $s_{ij} =\sin\theta_{ij}$ , $c_{ij} =\cos\theta_{ij}$ , and $\delta$ is the phase responsible for all CP-violating phenomena in flavor-changing processes in the SM. The angles $\theta_{ij}$ can be chosen to lie in the first quadrant, so $s_{ij}$ , $c_{ij}\ge 0$.
 The unitarity of the CKM matrix, ${\bf V}^{\dagger}{\bf V} ={\bf 1}$, is the modern way to search for physics beyond the SM. In case of this requirement is violated, it is an indication of the existence of a fourth quark family.

The origin of magnitudes and phases in the CKM matrix is intimately connected
with the origin of the quark masses themselves and family replication, in the SM; the physics for the quark masses is related to the Yukawa interactions and the Higgs mechanism. In the basis where the SM Lagrangian is gauge invariant, the charged currents are diagonal and after the electroweak symmetry breaking (EWSB) the quarks, leptons, Higgs fields  and gauge bosons acquire mass, but the mass matrices are not diagonal.

Rotating the up and down quark mass matrices to its diagonal form, rotates and mixes the quark fields, and we obtain  the corresponding mass eigenvalues; in this basis the charged currents are not diagonal,  we obtain the CKM matrix instead. 

T.D. Lee noticed that CP violation effects may arise after the EWSB in a model with more than one Higgs.  One should keep in mind, in particular, the possibility that there may be a close link between the Higgs sector and CP violation,  and one should be open to the possible appearance of SU(2) Higgs doublets  of non-standard representations such as singlets and doublets under a flavor symmetry group.  Revealing in this way that the full details of the underlying mechanism of electroweak symmetry breaking may be considerably more complex than in the Standard Model.

In the SM due to the form of the scalar potential, one component of the complex scalar field acquires a vacuum expectation value (vev), and the SU(2)$_L\times$ U(1)$_Y$ electroweak symmetry is spontaneously broken down to the U(1)$_{EM}$  gauge symmetry of electromagnetism. 

Hermiticity of the Lagrangian requires real SM scalar potential parameter, and the bosonic sector of the electroweak theory is CP-conserving.
As we know, the SM with only one Higgs, provides an extremely successful description of the observed electroweak phenomena. Nevertheless, there are a number of motivations to extend the Higgs sector of this model by adding a second and third complex SU(2) doublet of scalar fields.

The theoretical structure of the Standard Model   is constrained by  the flavor group $S(3)$ ($S(3)$SM). This model contains  three-Higgs doublet fields,  leading to numerous relations among Higgs masses and couplings. In particular, as in the case of the SM, in the normal minimum the tree-level $S(3)$SM Higgs sector is CP-conserving~\cite{EmmanuelCosta:2007zz,Beltran:2009zz}. 

The $S(3)$SM Lagrangian contains twelve  real scalar fields. After electroweak symmetry breaking, three Goldstone bosons ($G^{\pm}$ and $G^0$) are removed from the spectrum and provide the longitudinal modes of the massive $W^{\pm}$ and $Z^0$.  Nine physical Higgs particles remain: two charged Higgs pairs ($H^{\pm}$) and five neutral Higgs. If experimental data reveals the existence of a Higgs sector beyond that of the SM, it will be crucial to test whether the observed scalar spectrum is consistent with an $S(3)$SM interpretation.

In order to be general within this framework, one should allow for  CP-violating  arising from $S(3)$SM when confronting the data. Any observed relations among the  $S(3)$SM parameters would surely contribute to the search for a deeper theoretical understanding of the origin of CP arising from the electroweak symmetry breaking.

In this work we perform a detailed study of the spontaneous CP breaking conditions of  $S(3)$SM.  This model has been previously used to calculate the Higgs masses spectrum and mixings as well as trilinear Higgs self-couplings~\cite{Barradas-Guevara:2014yoa}, quark and lepton mixing~\cite{Caravaglios:2005gw,Canales:2013cga}, and flavour changing neutral currents (FCNC)~\cite{Kubo:2005sr,Mondragon:2007af} with success.

The model has three $S(3)$ flavoured Higgs fields, $H_1$, $H_2$ and $H_S$, which upon acquiring vev's break the electroweak symmetry.  Here, we examine the CP breaking minimization conditions, without explicitly breaking the flavour symmetry, although it might be spontaneously broken. $S(3)$SM has tree different stationary points, which can be
classified as Normal, Charge Breaking (CB) and Charge Parity Breaking
(CPB) minima, according to the vacuum expectation value of the three
Higgs fields~\cite{Beltran:2009zz}.

We describe the CPB scenarios of the model and give expressions
for the Higgs mass matrix in one CPB scenario.  A numerical computation of  the trilinear self-couplings  $\lambda_{H_i^0H_i^0H_i^0}, \,  i = 1,\cdots,5$ allows us to find out $H_4^0$, which is the Higgs candidate like to the SM one.

\section{ $S(3)$ Standard Model Fields \label{sec:potential}} 
In writing the Higgs potential, we have implicitly chosen a basis in the three-dimensional  flavor space of scalar fields. After EWSB, in the CP minimum, the $S(3)$SM scalar potential depends on real parameters and six Higgs vev's, for a total of sixteen degrees of freedom. 
The $S(3)$SM is governed by the choice of the Higgs potential and the Yukawa couplings of the three scalar to the three generations of quarks and leptons.  In the gauge basis $\Phi_a$ , $\Phi_b$ and $\Phi_c$ denote three complex hypercharge-one scalar fields: 
\begin{equation}
H_{s} = \frac{1}{\sqrt{3}}\Bigl(\Phi_{a} + \Phi_{b} +
\Phi_{c}\Bigr),\qquad \left(
\begin{array}{c} 
H_1\\
\\
H_2
\end{array} \right)
= \left(
\begin{array}{c}
 \frac{1}{\sqrt{2}}(\Phi_{a}-\Phi_{b})\\
 \\
\frac{1}{\sqrt{6}}(\Phi_{a}+\Phi_{b}-2\Phi_{c})
\end{array} \right).
\end{equation}

The quark, lepton and Higgs fields transform in the same way under the flavour symmetry group $S(3)$, all the fields have three species (flavours) and belong to a representation reducible to ${\bf 1_S}\oplus {\bf 2}$ of $S_{3}$. This implies that  $Q_{s}$ and $Q_{1,2}$ for left and right up or down quark fields, and  $L_{s}$ as well as $L_{1,2}$ for left and right charged or neutral lepton fields. 

\section {$S(3)$ Invariant Yukawa Lagrangian}
\noindent The $S(3)$SM is governed by the choice of the Higgs potential and the Yukawa couplings of the three scalar for the three generations of quarks and leptons.  Let $H_1$ , $H_2$  and $H_S$ denote three complex hypercharge-one, SU(2)$_L$ doublet scalar fields. The challenge is to develop a model that incorporates entirely or partially the CKM mechanism, in this direction, the Yukawa Lagrangian may be written as~\cite{Kubo:2003iw}
\begin{equation}
\begin{array}{rcl}
{\cal L}_{Y_D} &=&
- Y_1^d \overline{ Q}_I H_S d_{IR} - Y_3^d \overline{ Q}_3 H_S d_{3R} \\ \nonumber
& &-   Y^{d}_{2}[~ \overline{ Q}_{I} \kappa_{IJ} H_1  d_{JR}
+\overline{ Q}_{I} \eta_{IJ} H_2  d_{JR}~] \\ \nonumber
& &- Y^d_{4} \overline{ Q}_3 H_I  d_{IR} - Y^d_{5} \overline{ Q}_I H_I d_{3R} 
+H.c., \\
{\cal L}_{Y_U} &=&
-Y^u_1 \overline{ Q}_{I}(i \sigma_2) H_S^* u_{IR} 
-Y^u_3\overline{ Q}_3(i \sigma_2) H_S^* u_{3R} \\ \nonumber
&& -Y^{u}_{2}[~ \overline{ Q}_{I} \kappa_{IJ} 
(i \sigma_2)H_1^*  u_{JR} 
+ \eta  \overline{ Q}_{I} \eta_{IJ}(i \sigma_2) H_2^*  u_{JR}~] \\ \nonumber
&&-Y^u_{4} \overline{ Q}_{3} (i \sigma_2)H_I^* u_{IR} 
-Y^u_{5}\overline{ Q}_I (i \sigma_2)H_I^*  u_{3R} +H.c.,
\end{array}
\end{equation}
Singlets carry the index $s$ or $3$, and doublets carry indices $I, J = 1,2 $
\begin{equation}
\kappa = \left( 
\begin{array}{cc}
0 & 1 \\
1 & 0
\end{array} \right) 
\quad 
{\rm and} 
\quad
\eta = \left(
\begin{array}{cc}
1 & 0 \\
0 & -1
\end{array} \right) .
\end{equation}

The $S(3)$ lepton and neutrino invariant Yukawa Lagrangian may also  be written by substituting the down quark sector in terms of the charged leptons and up sector by neutral leptons. Furthermore, in this model the mass terms for the Majorana neutrinos may also be introduced,
\begin{equation}
 {\cal L}_{M} = -M_1
\nu_{IR}^T C \nu_{IR} -M_3 \nu_{3R}^T C \nu_{3R},\nonumber 
\end{equation}
where $C$  is the charged matrix.

\section {$S(3)$ Invariant Higgs Lagrangian}
The Lagrangian  ${\cal L}_{H}$ of the $S(3)$ extended Higgs sector incorporates three complex scalar SU(2) doublets fields. Such theory is  based on purely flavour symmetry grounds: in view of the family replication of the elementary fermion spectrum one can speculate that this flavour symmetry is the symmetry of the fundamental particles,  and an  analogous flavour symmetry principle might work for the Higgs sector too: 
\begin{equation}
{\cal L}_{\Phi_i} =\left[   D_\mu \Phi_S\right]^2+\left[D_\mu \Phi_1\right]^2
+\left[D_\mu \Phi_2\right]^2-V\left( \Phi_1,\Phi_2,\Phi_S \right),
\label{eq:one}
\end{equation}
where $D_\mu$ is the usual covariant derivative,
$ D_\mu=\left( \partial_\mu-\frac{i}{2}g_2{\tau_a}  {W_{\mu}^a}-\frac{i}{2}g_1B_\mu \right)$, with $g_1$ and $g_2$ standing for the U(1) and SU(2) coupling constants. 

The only terms that can possibly violate the CP symmetry are the Yukawa couplings and scalar potential. In the SM, the unique source of CP violation comes from the complex phases in the Yukawa couplings that are transferred to the CKM matrix~\cite{bib3} after EWSB. Within such context the possibility of (explicit) CP violation is intimately connected with the presence of a horizontal space: the quarks come in three identical families distinguished only by their masses.

As we know, explicit and spontaneous CP violation can arise in the scalar potential in addition to EWSB~\cite{Nishi:2006tg}. Electroweak symmetry breaking arises if the minimum of the scalar potential occurs for nonzero expectation values of the scalar fields, 
\begin{equation} \label{eq:doubletshiggs}
H_1=\left( 
\begin{array}{c}
\phi_1+i\phi_4\\
\phi_7+i\phi_{10}
\end{array} \right),\
H_2=\left( 
\begin{array}{c}
\phi_2+i\phi_5\\
\phi_8+i\phi_{11}
\end{array} \right),\
H_S=\left( 
\begin{array}{c}
\phi_3+i\phi_6\\
\phi_9+i\phi_{12}
\end{array} \right).
\end{equation}
The subscript $s$ is the flavour index for the Higgs field singlet under $S(3)$. $H_i$ with $i=1,2$ are the components of the $S(3)$ doublet field. It is always possible to write the $S(3)$ scalar potential in the following form
\begin{equation} \label{eq:potential2}
V({\bf X} )={\bf A}^T{\bf X}+\frac{1}{2}{\bf X}^T{\bf B}{\bf X},
\end{equation}
with the vector  ${\bf X}^T=\left( x_1, x_2 ,x_3, \dots,x_9 \right)$ given by nine real quadratic forms $x_i$
\begin{equation}\label{eq:xveccomp}
\begin{array}{ccc}
x_1=H^\dagger_1 H_1,&
x_4= {\cal R}\left(H^\dagger_1 H_2\right),&
x_7= {\cal I}\left(H^\dagger_1 H_2\right),\cr
x_2=H^\dagger_2 H_2,&
x_5= {\cal R}\left(H^\dagger_1 H_S\right),&
x_8= {\cal I}\left(H^\dagger_1 H_S\right),\cr
x_3=H^\dagger_S H_S ,&
x_6= {\cal R}\left(H^\dagger_2 H_S\right),&
x_9={\cal I}\left(H^\dagger_2 H_S\right) .
\end{array}
\end{equation}
${\bf A}$ is a mass parameter vector,
\begin{eqnarray}\label{eq:amatrix}
 {\bf A}^T=\left(\mu^2_1,\mu^2_1,\mu^2_0,0,0,0,0,0,0 \right)  
 \label{appppb}
\end{eqnarray}
and ${\bf B}$  is a $9\times9$ real parameter symmetric matrix
%
\begin{equation} \label{eq:bmatrix}
{\bf B}= \left( \begin{array}{ccccccccc}
2(c+g) & 2(c-g) &b&0&0&2e&0&0&0\cr
2(c-g) & 2(c+g) & b&0&0&-2e&0&0&0\cr
b&b&2a&0&0&0&0&0&0\cr
0&0&0&8g&4e&0&0&0&0\cr
0&0&0&4e&2(f+2h)&0&0&0&0\cr
2e&-2e&0&0&0&2(f+2h)&0&0&0\cr
0&0&0&0&0&0&-8d&0&0\cr
0&0&0&0&0&0&0&2(f-2h)&0\cr
0&0&0&0&0&0&0&0&2(f-2h)
\end{array}
\right).
\end{equation}
%
$\mu_0, \mu_1,$ and $a,b,c,d,e,f,g,h$ are real parameters. The $S(3)$ invariant Higgs potential in Eq.~(\ref{eq:potential2})  has a CPB  minimum at  
\begin{equation}\label{eq:minimcpviolation123}
\begin{array}{c}
\phi_7=v_1,\ \phi_8=v_2,\ \phi_9=v_3, \
\phi_{10}= \gamma_1,\ \phi_{11}=\gamma_2,\ \phi_{12}=\gamma_3,\qquad \hbox{and other cases} \qquad \phi_i=0,
\end{array}
\end{equation}
where $v_{1,2,3}$ and $\gamma_{1,2,3}$ are real numbers.    
Here we present the analysis in the case where CPB comes from the Higgs singlet $H_S$
this vacuum solution preserves the U(1)$_{EM}$ symmetry; it corresponds to a local minimum of the potential if its parameters are such that the physical Higgs squared-masses are non-negative.

\section{Minimum Conditions \label{sec:mincond}}
In this section, we present the minimum conditions and the parameter space analysis for the considered CPB scenario with $\gamma_3 \neq 0$ and $\gamma_1=\gamma_2=0$. Electroweak symmetry breaking arises if the minimum of the scalar potential occurs for nonzero expectation values of the scalar fields.
\begin{equation}
\langle H_i \rangle = \displaystyle\frac{1}{\sqrt{2}} \left( \begin{array}{c} 0 \\ v_i + i \gamma_i \end{array}\right) \, \qquad
i = 1, 2, 3,
\end{equation}
where $v_i,\, \gamma_i \in \mathbb{R}$e. The conditions for extrema of the $S(3)$ scalar potential, considering the CPB conditions of the scalar potential are 
\begin{eqnarray}
M_7({\gamma_3}) &=&2 v_1 \left(\gamma _3^2 (b+f-2 h)+v_3^2 (b+f+2 h)+2 v_1^2 (c+g)+2 v_2^2 (c+g)+6 e v_2
   v_3+\mu _1^2\right)\label{eq:condmin3a},
\\
M_8({\gamma_3}) &=& 2 \left(v_2 \left(\gamma _3^2 (b+f-2 h)+v_3^2 (b+f+2 h)+2 v_2^2 (c+g)-3 e v_3 v_2+\mu
   _1^2\right) +v_1^2 \left(2 v_2 (c+g)+3 e v_3\right)\right)\label{eq:condmin3b},
\\
M_9({\gamma_3}) &=& 2 \left(v_3 \left(2 a \left(\gamma _3^2+v_3^2\right)+\mu _0^2\right)+v_1^2 \left(v_3
   (b+f+2 h)+3 e v_2\right)+v_3 v_2^2 (b+f+2 h)-e v_2^3\right) \label{eq:condmin3c},
\\
M_{10}(\gamma_3) &=&  4 \gamma _3 v_1 \left(e v_2+2 h v_3\right)\label{eq:condmin3d},
\\
M_{11}(\gamma_3) &=& 2 \gamma _3 \left(v_2 \left(4 h v_3-e v_2\right)+e v_1^2\right)\label{eq:condmin3e},
\\
M_{12}(\gamma_3) &=& 2 \gamma _3 \left(2 a \left(\gamma _3^2+v_3^2\right)+\left(v_1^2+v_2^2\right) (b+f-2
   h)+\mu _0^2\right).\label{eq:condmin3f}
\end{eqnarray}
Here,
$$
M_{i}(\gamma_3) = \left.\displaystyle\frac{\partial V(\phi_1,\phi_2,\cdots,\phi_{12})}{\partial\phi_i}\right|_{\rm min} = 0 
$$
with minimum in
\begin{equation}\label{eq:minimcpviolation}
{\rm min} = \bigl\{\ \phi_7=v_1, \ \phi_8=v_2, \ \phi_9=v_3, \ \phi_{12} = \gamma_3, \quad {\rm and\ other\ cases} \ \phi_i = 0 \ \bigr\}.
\end{equation}
From Eqs.~(\ref{eq:condmin3a}) and~(\ref{eq:condmin3b}), we have
$$
v_2M_7 - v_1M_8 = 0.
$$
Then,
\begin{equation}\label{eq:v1}
v_1= \sqrt{3} v_2 .
\end{equation}
Hence, from Eq.~(\ref{eq:condmin3d}), where we take this equal to zero, so that
\begin{equation}
h =  -\displaystyle\frac{e v_2}{2v_3},
\end{equation}
with these values for $v_1$ and $h$  from Eqs.~(\ref{eq:condmin3a}),~(\ref{eq:condmin3f}), we obtain 
%
\begin{eqnarray}\label{eq:valmu0c3}
\mu_0^2 &=&
-\displaystyle{\frac{a v_3 \left(\gamma _3^2+v_3^2\right)+2 v_3 v_2^2 (b+f)+2 e
   v_2^3}{v_3}},\\
\mu_1^2 &=&
-\displaystyle{\frac{v_3 (b+f) \left(\gamma _3^2+v_3^2\right)+8 v_3 v_2^2 (c+g)+e v_2 \left(\gamma
   _3^2+5 v_3^2\right)}{2v_3}} .
   \label{eq:valmu1c3}
   \end{eqnarray}
%

The minimum constraints are the conditions ensuring that the extremum is a minimum for all directions in $\phi$'s space, except for the direction of the Goldstone modes. It is realized when the squared-masses of the nine physical Higgses are all positive.
The tree-level amplitudes for the scattering of longitudinal gauge bosons at high energy can be related via the equivalence theorem to the corresponding amplitudes in which the longitudinal gauge bosons are replaced by Goldstone bosons. 
 
We have adopted for convenience $v_i \in \mathbb{R}$e ($i=1,2,3$). In a standard notation, we can write them as 
\begin{equation}
\label{eq:vevsrel}
\begin{array}{rcl}
v_2&=&v( \sin{\omega_3}\sin{\omega_{CP}})/2, \\ 
v_3&=&v \cos{\omega_3}\sin{\omega_{CP}}, \\
\gamma_3&=&v \cos{\omega_{CP}},
\end{array}
\end{equation}
where $\omega_3$, $\omega_{CP}$ are two free mixing parameter angles. The scalar potential with spontaneous CP-breaking coming from the Higgs singlet $H_S$ has a minimum in Eq.~(\ref{eq:minimcpviolation}), which have to satisfy the constraint
\begin{equation}
v= \left(v_1^2+v_2^2 + v_3^2+ \gamma_3^2\right)^{1/2}= 246 \, \textrm{GeV}.\label{eq:constraint}
\end{equation}
We assume Higgs vev's as free parameters subject to the constraint~(\ref{eq:constraint}), the potential parameters in Eq.~(\ref{eq:potential2}), specifically the mass parameters $\mu_0^2$ and $\mu_1^2$, may be written in terms of the vev's. 
In the Standard Model (SM), the unique source of CP violation comes from the complex phase in the Cabibbo-Kobayashi-Maskawa (CKM) matrix which may be transferred from the Yukawa couplings after spontaneous electroweak symmetry breaking (EWSB). For explicit CP violation the Yukawa couplings may all be complex.
The possibility of (spontaneus) CP violation is intimately connected with the presence of a horizontal space: the quarks come in three identical families distinguished only by their masses as well as the leptons and Higgs fields as it is assumed in the $S(3)$SM. In particular  for real Yukawa couplings the corresponding Yukawa Lagrangian is given in Ref.~\cite{Barradas-Guevara:2014yoa,Kubo:2003iw}. From this, we can express the fermionic mass  matrix ${\bf M}_f$ including spontaneous CP violation ($\gamma_3 \neq 0)$ as
\begin{equation}
{\bf M}_f=\left( 
\begin{array}{ccc}
m_{1}^{CP}+m_{6} & m_{2}^{{}} & m_{5}^{{}} \\ 
m_{2}^{{}} & m_{1}^{CP}-m_{6} & m_{8} \\ 
m_{4}^{{}} & m_{7} & m_{3}^{CP}
\end{array}
\right) ,
\end{equation}
where the CP breaking phase is 
\begin{eqnarray}
m_{1}^{CP} &=&m_{1} -Y_{1}^{f}\left( i\gamma_3 \right), \\
m_{3}^{CP} &=&m_{3} -Y_{3}^{f}\left(i\gamma_3 \right). 
\end{eqnarray}
The remaining $m_{i} \, (i=1,2,\cdots,8$), are the same as in the CP conserving minimum~\cite{Kubo:2003iw}. Thus, the EWSB mechanism provides a source for CP violation in the fermion sector contributing with a CP phase in the quark and lepton mixing matrices. In next section we compute the Higgs mass matrix in the CPB minimum Eq.~(\ref{eq:minimcpviolation}), the corresponding Higgs mass eigenvalues are given as well. 

\section{Higgs Masses}
If we want to know the Higgs mass matrix $\mathcal{M}^2_H$ , it is necessary to compute the second derivatives of the Higgs potencial, Eq.~(\ref{eq:potential2}), at the CPB minimum conditions.  
\begin{equation}
(\mathcal{M}^2_H)_{ij} = \displaystyle\left.\displaystyle\frac{1}{2}\displaystyle\frac{\partial^2 V}{\partial\phi_i\partial\phi_j}\right|_{\rm min}~~{\rm with}~~i,j={1,2,\cdots.12},
\end{equation}
For three Higgs fields $H_1$, $H_2$ and $H_S$, Eq.~(\ref{eq:doubletshiggs}), the corresponding $\mathcal{M}^2_H$ is a 12 $\times$ 12 real matrix. The Higgs mass matrix is block diagonal with $6\times 6$ symmetric and Hermitian sub-matrices corresponding to the electrically charged ${\bf M}_{C,\gamma}^2$ and the neutral ${\bf M}_{N,\gamma}^2$ bosons mass matrix respectively. The charged Higgs mass matrix is given by 
\begin{equation}
{\bf M}^2_{C, \gamma} = \left(
\begin{array}{cc}
{\bf {M}^2_C}_{11}(\gamma) & {\bf {M}^2_C}_{12}(\gamma)  \\ [.4cm]
-{\bf {M}^2_C}_{12}(\gamma) & {\bf {M}^2_C}_{11}(\gamma) 
\end{array} \right), \label{eq:matrizcargada}
\end{equation}
the mass sub-matrices for charged Higgs in Eq.~(\ref{eq:matrizcargada}) are given by
\begin{equation}
{\bf {M_C}}^2_{11}(\gamma_3) = 
\left(
\begin{array}{ccc}
 -2 g v_2^2+\displaystyle\frac{e
   \left(3 v_3^2+\gamma
   _3^2\right) v_2}{2v_3}+\displaystyle\frac{f 
   \left(v_3^2+\gamma
   _3^2\right)}{2} &
   \sqrt{3} v_2 \left(2 g
   v_2+e v_3\right) &
   \displaystyle\frac{1}{2} \sqrt{3} v_2
   \left(e v_2+f v_3\right) \\
  \sqrt{3} v_2 \left(2 g
   v_2+e v_3\right) & -6 g
    v_2^2+\displaystyle\frac{e \left(7
   v_3^2+\gamma _3^2\right)
   v_2}{2v_3}+\displaystyle\frac{f 
   \left(v_3^2+\gamma
   _3^2\right)}{2} &
   \displaystyle\frac{1}{2} v_2 \left(e
   v_2+f v_3\right) \\
   \displaystyle\frac{1}{2} \sqrt{3} v_2
   \left(e v_2+f v_3\right) &
   \displaystyle\frac{1}{2} v_2 \left(e
   v_2+f v_3\right) & -\displaystyle\frac{2
   v_2^2 \left(e v_2+f
   v_3\right)}{v_3} \\
\end{array}
\right), \label{eq:mc11g3}
\end{equation}
\begin{equation}
{\bf {M_C}}^2_{12}(\gamma_3) =
\left(
\begin{array}{ccc}
 0 & 0 & \displaystyle\frac{1}{2} \sqrt{3}
   v_2 \left(f+\displaystyle\frac{e
   v_2}{v_3}\right) \gamma _3
   \\
 0 & 0 & \displaystyle\frac{1}{2} v_2
   \left(f+\displaystyle\frac{e
   v_2}{v_3}\right) \gamma _3
   \\
 -\displaystyle\frac{1}{2} \sqrt{3} v_2
   \left(f+\displaystyle\frac{e
   v_2}{v_3}\right) \gamma _3
   & -\displaystyle\frac{1}{2} v_2
   \left(f+\displaystyle\frac{e
   v_2}{v_3}\right) \gamma _3
   & 0 \\
\end{array}
\right). \label{eq:mc21g3}
\end{equation}

Now, we substituted the Eqs.~(\ref{eq:mc11g3}) and~(\ref{eq:mc21g3}) in~(\ref{eq:matrizcargada}) and diagonalize the resulting matrix. The  eigenvalues for the charged Higgs fields are degenerated as follows 
\begin{equation}
\left\{0,-\displaystyle\frac{\left(\gamma
   _3^2+4 v_2^2+v_3^2\right)
   \left(e v_2+f v_3\right)}{2
   v_3},-\displaystyle\frac{e v_2
   \left(\gamma _3^2+9
   v_3^2\right)+f v_3
   \left(\gamma
   _3^2+v_3^2\right)+16 g v_3
   v_2^2}{2 v_3}\right\},
\end{equation}
That is, there are two massless and four non-massless Higgs fields:
\begin{equation}\label{eq:masscharged}
\begin{array}{l}
M_{H_1^\pm} =  -\displaystyle\frac{v^2
   \left(e v_2+f v_3\right)}{2
   v_3} \\
M_{H_2^\pm} =  -\displaystyle\frac{e v_2
   \left(\gamma _3^2+9
   v_3^2\right)+f v_3
   \left(\gamma
   _3^2+v_3^2\right)+16 g v_3
   v_2^2}{2 v_3}
\end{array}
\end{equation}

The neutral Higgs mass matrix has the form
\begin{equation}
{\bf M}^2_{N,\gamma} = \left(
\begin{array}{cc}
{\bf {M}^2_N}_{11}(\gamma) & {\bf {M}^2_N}_{12}(\gamma) \\[.4cm] 
{\bf {M}^2_N}_{12}^{\rm T}(\gamma) & {\bf {M}^2_N}_{22}(\gamma) 
\end{array} \right), \label{eq:matrizneutra}
\end{equation}
here ${\bf {M}^2_N}_{12}^{\rm T}(\gamma)=\left({\bf {M}^2_N}_{12}(\gamma)\right)^T$ and the  neutral $3\times3$ Higgs sub-matrices are given by
\begin{equation}
{\bf {M_N}}^2_{11}(\gamma_3) =
\left(
\begin{array}{ccc}
 6 (c+g) v_2^2 & \sqrt{3} v_2
   \left(2 (c+g) v_2+3 e
   v_3\right) & \sqrt{3} v_2
   \left(2 e v_2+(b+f)
   v_3\right) \\
\sqrt{3} v_2
   \left(2 (c+g) v_2+3 e
   v_3\right) & 2 v_2
   \left((c+g) v_2-3 e
   v_3\right) & v_2 \left(2 e
   v_2+(b+f) v_3\right) \\
 \sqrt{3} v_2
   \left(2 e v_2+(b+f)
   v_3\right) & v_2 \left(2 e
   v_2+(b+f) v_3\right) & 2 a
   v_3^2-\displaystyle\frac{4 e v_2^3}{v_3}
   \\
\end{array}
\right), \label{eq:mn11g3}
\end{equation}

\begin{equation}
{\bf {M_N}}^2_{12}(\gamma_3) =
\left(
\begin{array}{ccc}
 0 & \sqrt{3} e v_2 \gamma _3
   & \displaystyle\frac{\sqrt{3} v_2
   \left(e v_2+(b+f)
   v_3\right) \gamma _3}{v_3}
   \\
 \sqrt{3} e v_2 \gamma _3 & -2
   e v_2 \gamma _3 & \displaystyle\frac{v_2
   \left(e v_2+(b+f)
   v_3\right) \gamma _3}{v_3}
   \\
 -\displaystyle\frac{\sqrt{3} e v_2^2
   \gamma _3}{v_3} & -\displaystyle\frac{e
   v_2^2 \gamma _3}{v_3} & 2 a
   v_3 \gamma _3 \\
\end{array}
\right), \label{eq:mn12g3}
\end{equation}
\begin{equation}
{\bf {M_N}}^2_{22}(\gamma_3) =
\left(
\begin{array}{ccc}
 -\displaystyle\frac{v_2 \left(2 (d+g) v_2
   v_3+e \left(v_3^2+\gamma
   _3^2\right)\right)}{v_3} &
   \sqrt{3} v_2 \left(2 (d+g)
   v_2+e v_3\right) & 0 \\
\sqrt{3} v_2 \left(2 (d+g)
   v_2+e v_3\right) &
   -\displaystyle\frac{v_2 \left(6 (d+g)
   v_2 v_3+e \left(3
   v_3^2+\gamma
   _3^2\right)\right)}{v_3} &
   0 \\
 0 & 0 & 2 a \gamma _3^2 \\
\end{array}
\right). \label{eq:mn22g3}
\end{equation}

Hence, we computed the neutral Higgs mass matrix given in Eq.~(\ref{eq:matrizneutra}) with Eqs.~(\ref{eq:mn11g3}),~(\ref{eq:mn12g3}) and~(\ref{eq:mn22g3}). Diagonalizing the resulting neutral Higgs mass matrix we obtain: one zero and five non zero eigenvalues, $M_{H_1^0}, \cdots, M_{H_5^0}$ . In accordance with the SM there is one neutral massless eigenvalue. Then, in the $S(3)$SM with spontaneous CPB coming from the Higgs singlet $H_S$ there are three Goldstone bosons, one neutral and one charged pair. We also obtained nine massive Higgs fields four electrically charged, with degenerated masses, two by two, and five neutral. The neutral Higgs mass eigenvalues are shown in Figure~\ref{fig:masses}. That is, we have got a  physically acceptable final result, because in this spontaneous CPB minimum we have obtained three  Goldstone bosons which can give mass to vector bosons $W^{\pm}$ and $Z^0$ with a massless photon and nine physical Higgs fields, at least one neutral Higgs could have a mass of $125.7\pm0.4$ GeV, and the remaining eight additional Higgs states are candidates for new particles. This scenario provides a strong motivation to extend the analysis  to CPB phenomenology arising from spontaneous electroweak symmetry breaking. 
We denote the masses of these charged Higgses as $M_{C_i}$ and $M_{H_j^0}$ for the neutral Higgs masses, where $i=1,2$ and $j=1,\cdots,5$.  
In the following section,  we analyze the Higgs masses in the spontaneous CPB minimum where $v_1=\sqrt{3}v_2$.

\section{$S(3)$SM Parameter Space \label{sec:parameterspace}}
\begin{figure}[ht] 
\centering 
\includegraphics[scale=1.1]{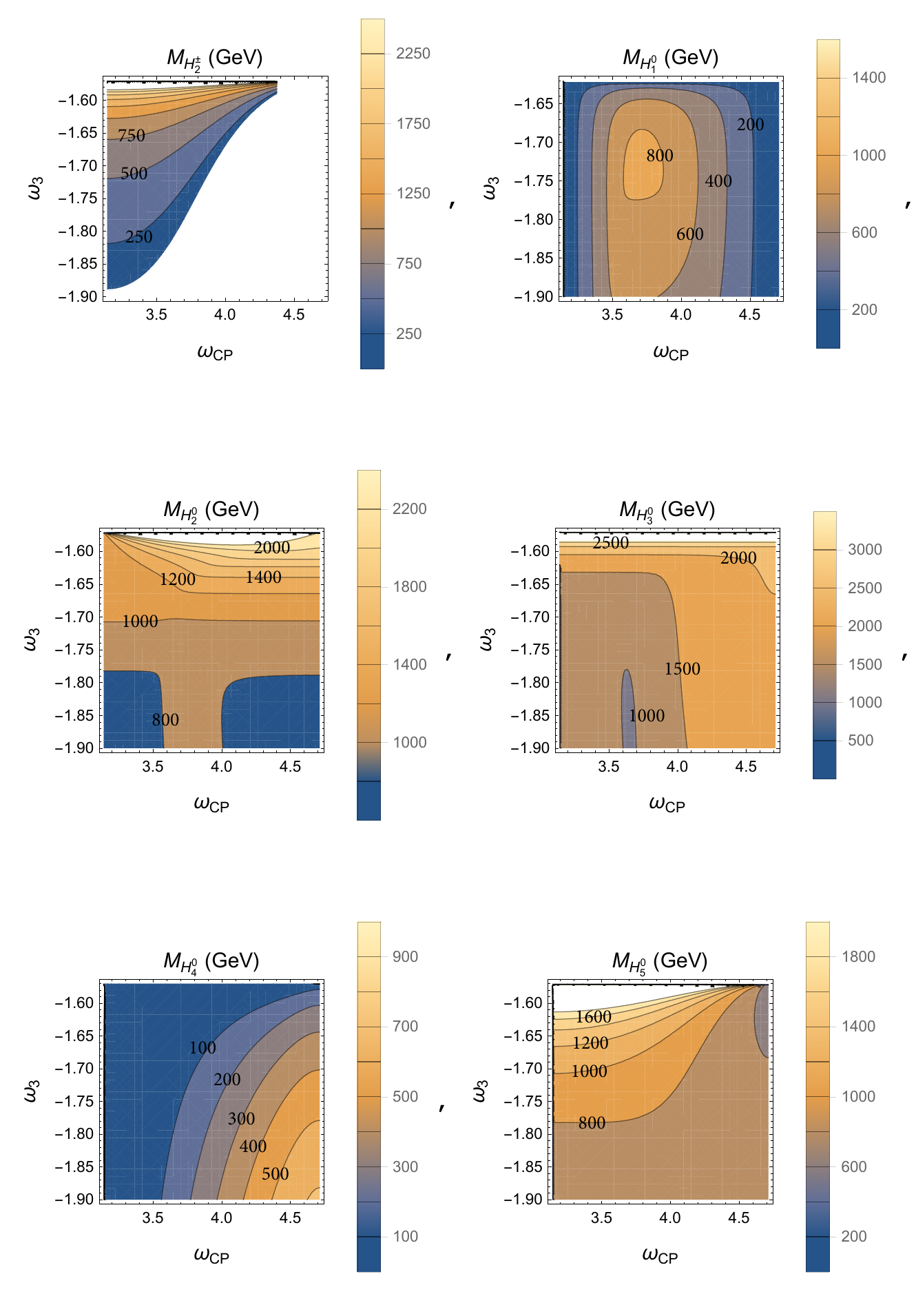} 
\caption{The Higgs masses $M_{H_2^\pm}$ and $M_{H_i^0}\, (i=1,\cdots,5)$ for $a=10$, $b=1$, $c= 30$, $d=-6$, $e=-4.54$, $f=6.9$, $g=3.5$, as a function of the angles $\omega_3$, $\omega_{CP}$, where $\pi \leqq \omega_{CP} \leqq 3\pi/2$ and $-1.9 \leqq \omega_{3} \leqq -1.57$. In this region, the Higgses ${H_j^0}$ for $j=1,4$ are candidates for a SM like Higgs with mass values at $125.7\pm{10}$ GeV while $M_{H_j^0}$ for $j=2,3,5$ will never reach this value. The $M_{H_1^\pm}$ is not shown, and the lower and upper levels of $\omega_3$, $\omega_{CP}$ are set from the masses charged Higgs.}  
\label{fig:masses}
\end{figure}

In this section, we explore $\gamma_3$ parameter space regions where the model is consistent.  In order to do that, we need to map out regions in the space $(v_3, \gamma_3)$  where the parameter matrix ${\bf B}$, Eq.~(\ref{eq:bmatrix}), is positive definite, ensuring that we have a CPB minimum point in the $S(3)$ Higgs potential.  We assume that the set of quadratic couplings have values where the Higgs  masses are $\lesssim 2$ TeV, Table~\ref{table:two}. Since $M_{H_1^\pm}$ only depend of $\omega_3$, Eq.~(\ref{eq:masscharged}), then $-1.9 \leq \omega_{3} \leq -1.57$, and from the values for masses de Higgses  of Table~\ref{table:two}; our numerical analysis is performed using 
\begin{equation}\label{eq:numericalpar}
\begin{array}{c}
\big\{\ a = 10,\ b = 1,\ c = 30,\ d = -6,\ e =  -4.54,\ f  =  6.9,\ g  = 3.5,\ h < f/2 \ \big\}\, ;
\end{array}
\end{equation}
with such values, the matrix ${\bf B}$, Eq.~(\ref{eq:bmatrix}), is positive definite, and the Higgs mass eigenvalues are positive too. Also, these parameter values allowed us to have the mass of the lightest neutral Higgs $H^0_4$  to be less than $160$ GeV. We can see in Figure~\ref{fig:masses} the behavior of the masses with respect to the free parameters $(\omega_3, \omega_{CP})$. From Eqs.~(\ref{eq:constraint}) and~(\ref{eq:v1}), we have got Eq.~(\ref{eq:vevsrel}).
\begin{table}[ht]
\centering 
\begin{tabular}{c c c c c c c} 
\hline\hline 
$M_{H_1^\pm}$ & $M_{H_2^\pm}$ & $M_{H_1^0}$ & $M_{H_2^0}$ & $M_{H_3^0}$ & $M_{H_4^0}$&$M_{H_5^0}$ \\ [0.5ex] 
\hline 
 0-1400& 100-900 & 100-1100 & 500-955 & 974-1872 & 100-634 & 634-737  \\ [1ex] 
\hline 
\end{tabular}
\caption{Charged and neutral Higgs masses (GeV).} 
\label{table:two} 
\end{table}

The smaller Higgs masses are $M_{H_{1}^0}$ and  $M_{H_{4}^0}$ while bigger Higgs mass values are allowed for $M_{H_{2,3,5}^0}$  without reaching values greater than 
2 TeV. Thus, we evaluate the parameter space and found out values in the range $125.7\pm0.4$ GeV for the Higgs $H_{1,4}^0$, and the  remaining one $H_{2,3,5}^0$ is decoupled from this range, with a mass value running  from 300 GeV to 1800 GeV.

The explored ranges for $\omega_3$ and $\omega_{CP}$ are obtained by considering that $\mu_0^2<0$, Eq.~(\ref{eq:valmu0c3}), and $\mu_1^2<0$, Eq.~(\ref{eq:valmu1c3}). As we remarked before, with the parameter values given in Eq.~(\ref{eq:numericalpar}), the matrix ${\bf B}$, Eq.~(\ref{eq:bmatrix}), is positive definite as well as the CPB Higgs mass eigenvalues. As expected, the explored $\omega_3$ and $ \omega_
{CP}$ parameter range allows heavy neutral Higgses  whose masses are greater than 450 GeV, as you can see from Figure~\ref{fig:masses}.
Traditionally, in the Higgs potential~(\ref{eq:potential2}) the  quadratic  ($\mu_0^2$, $\mu_1^2$) and  quartic parameters ($a,b,\cdots,h$) determine the masses of the neutral and charged Higgses. Otherwise, and this is the approach followed here, one can take the free parameters $\omega_3$, $ \omega_ {CP}$, as input, and determine the parameters of the potential as derived quantities, but, some choices of input will lead to physically acceptable masses, others will not.

\begin{figure}[h] 
\centering 
\includegraphics[scale=1.1]{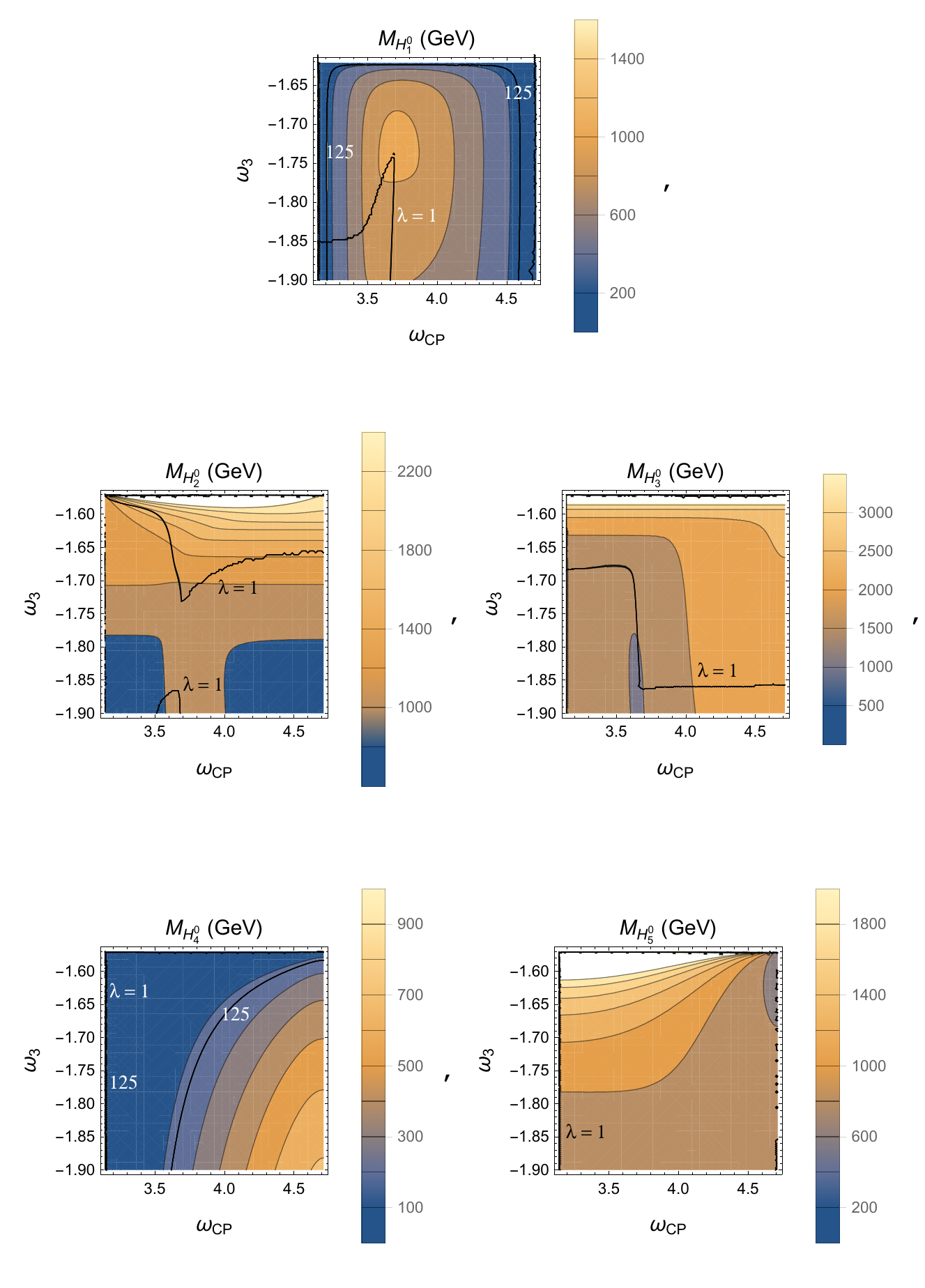}
\caption{The Higgs trilinear self-couplings $\tilde{\lambda}_{H_i^0H_i^0H_i^0} \, (i=1,\cdots,5)$, in the space $(\omega_3, \omega_{CP} )$, where $-1.90 \leqq \omega_3 \leqq -1.60$ and $\pi \leqq \omega_{CP} \leqq 3\pi/2$. Where the intersection is shown in masses $M_{H_i^0}$  with $\tilde{\lambda}_{H_i^0H_i^0H_i^0} \sim 1$. The coincidence is given to $M_{H_4^0} \sim 125$ GeV and  $\tilde{\lambda}_{H_4^0H_4^0H_4^0}\sim 1$, set $\omega_{CP} \sim 3.1$.}  \label{fig:lambdas}
\end{figure}

The potential (\ref{eq:potential2}) is attractive as one $S(3)$ extension of the SM that admits spontaneous CP violation. This is an interesting possibility, since it will become possible to severely constrain or even measure it. From this potential, we can derive as a function of the CP violation parameter $\gamma_3$ the Higgses trilinear self-couplings~\cite{Barradas-Guevara:2014yoa}, which are shown in Figure~\ref{fig:lambdas}.

In Figure~\ref{fig:masses}, the neutral Higgs masses  with respect to the parameters $\omega_ {3}$ and $\omega_ {CP} $ are shown, in which CP violation comes from the singlet $H_S$, Eq.~(\ref{eq:doubletshiggs}). It can be observed a light Higgs with $ M_ {H_ {2}^{0 }}$, $ M_ {H_ {4}^{0 }}$, $M_ {H_ {5}^{0}} < 960 \, \textrm {GeV} $, while the others are  $<1800 \, \textrm {GeV}$. Further, one can see that  five Higgses find a region in the parameter space such that they reach the values of the masses of $125.0\pm {0.01}$ GeV, see Figure~\ref{fig:lambdas}. One can be seen from Figure~\ref{fig:lambdas} that each neutral Higgs acquires mass values around 125 GeV  in different parameter space regions ($\omega_3$, $\omega_{CP}$).  Then, the computation of the self-couplings  allows us to identify a Higgs like the one in the SM. We have to look for  parameter space regions ($\omega_3$, $\omega_{CP}$) such that they simultaneously fit the Higgs mass and trilinear self-coupling for values as in the SM.
\begin{equation}
 \tilde{\lambda}_{H_i^0H_i^0H_i^0} \equiv \lambda_{H_i^0H_i^0H_i^0}/ \lambda_ {h ^0h^0h^0}^{SM}, \quad \lambda_ {h ^0h^0h^0}^{SM}=\displaystyle{\frac{3 M_{h^0}^{2}}{v}}.
  \label{eq:lamtilde}
 \end{equation}
Figure~\ref{fig:lambdas} show   
$ \tilde{\lambda}_{H_i^0H_i^0H_i^0}, \,  i = 1,\cdots,5$, that is,  in this model we can identify three possible  trilinear self-couplings candidates like to the one in the SM. Comparing with Figure~\ref{fig:lambdas}, we found that just $H_4^0$ is the only SM Higgs candidate in the range 110-140 GeV, with $\tilde{\lambda}_{H_4^0H_4^0H_4^0} \sim 1$.

%

\section{Conclusions}
\noindent We studied spontaneous CP violation in the most Higgs potential under the non-Abelian flavour symmetry S(3) of the extend SM. Introducing three SU(2)${}_L$ Higgs doublet fields in the theory, 
with twelve real fields.  A define structure of the CP violation is obtained which permits the calculation of mass for charge and neutral Higgs boson  and
we found a parameter space region where the minimum of the potential defines a CPB ground state. 
 A further reduction of free parameters is achieved in the sector scalar, by introducing a mixing angles of the critical points of Higgs potential. It can be seen in Figure 2, the Higgs masses in a region of space ($\omega_3, \omega_{CP}$). In this range window, $-1.9 < \omega_3 <$ -1.57 and $\pi < \omega_{CP} < 3\pi/2$, $M_{H_{1,\cdots, 5}^0}$ take values smaller than 2 TeV in agreement observations at this time.
 The Higgs masses and the trilinear self-coupling are computed in terms only two free parameters. 
 We have shown the Higgs masses and trilinear self-coupling for an allowed parameter set. In this case, the trilinear self-couplings analysis confirms our hypothesis: one can have CP violation resulting from the neutral Higgs sector with a Higgs mass and trilinear self coupling in accordance with the Standard Model one. Also, we have found a CPN ground state where the CPB comes from the $S(3)$ Higgs singlet. Furthermore, we also computed the Higgs masses and Higgs trilinear self-coupling $\tilde{\lambda}_{H_i^0H_i^0H_i^0}$, $i = 1, \cdots, 5$, in terms $\omega_3, \omega_{CP}$. Particularly when CP violation comes from the $S(3)$ singlet $H_S$, one observed $H_4^0$ as a possible candidate like the SM Higgs.

\section{ACKNOWLEDGMENTS}
This work has been partially supported  by \textit{CONACYT-SNI (M\'exico)}. ERJ acknowledges the financial support received from FEC and \textit{PROFOCIE (M\'exico)}.




\end{document}